\documentclass[12p.,twocolumn,aps]{revtex4-1}
\usepackage{amsmath}
\usepackage{amsfonts}
\usepackage{amssymb}
\usepackage{graphicx}
\usepackage{dcolumn}
\usepackage{bm}
\usepackage[sort&compress]{natbib}
\newcommand{\cmc}{\,\mathrm{cm}^{-3}}

\newcommand{\wcm}{\,\mathrm{W/cm}^2}
\newcommand{\mic}{\,\mu\mathrm{m}}
\newcommand{\mev}{\,\mathrm{MeV}}
\newcommand{\fs}{\,\mathrm{fs}}

{\bibpunct{}{}{,}{s}{}{}}

\begin{document}

\title{Laser Plasma Accelerators}
\author{V. Malka}
\affiliation{%
Laboratoire d'Optique Appliqu\'ee, ENSTA-ParisTech, CNRS, Ecole Polytechnique, UMR 7639, 91761 Palaiseau, France\\
}%

\date{\today}

\begin{abstract}
Research activities on laser plasma accelerators are paved by many significant breakthroughs. This review article provides an opportunity to show the incredible evolution of this field of research which has, in record time, allowed physicists to produce high quality electron beams at the GeV level using compact laser systems. I will show the scientific path that led us to explore different injection schemes and to produce stable, high peak current and high quality electron beams with control of the charge, of the relative energy spread, and of the electron energy. 
\end{abstract}

\maketitle

\section{Introduction}
The discovery of supra conductivity at a 4 Kelvin temperature in mercury cooled with liquid helium, made by K. Onnes in 1911, has opened a very active field of research. After decades of effort, researchers have been able to produce new superconductors working at ``high" temperature. Even if there remains open fundamental questions, mastering superconductivity has allowed the discovery of tremendous applications in magnetic resonance imaging (MRI) for medicine, in transportation (magnetic suspension trains), and in modern accelerators (superconducting cavities). In parallel, since the first 1.26 MeV Hg ion beam made by E. O. Lawrence, accelerators have gained in efficiency and in performance. With a market of more than 3 Billions dollars per year, accelerators are used today in many fields such as cancer therapy, ion implantation, electron cutting and melting, non destructive inspection, etc...For fundamental research, the most energetic machines (those that deliver particle beams with energies greater than 1 GeV represent only 1\% of the total number of accelerators) have been developed, for example for producing intense X rays beams in the free electron laser scheme for the study of ultra fast phenomena of interest for example in biology to follow the DNA structure evolution, or in material science to follow evolution of molecules or of crystal structures. Higher energies accelerators are crucial to answer to important questions regarding the origin of the universe, of the dark energy, of the number of space dimension, etc...The larger one, the Large Hadron Collider, is expected, for example, to reveal very soon properties of the Higgs boson. 

\begin{figure}[htbp]
\includegraphics[width=8.5cm]{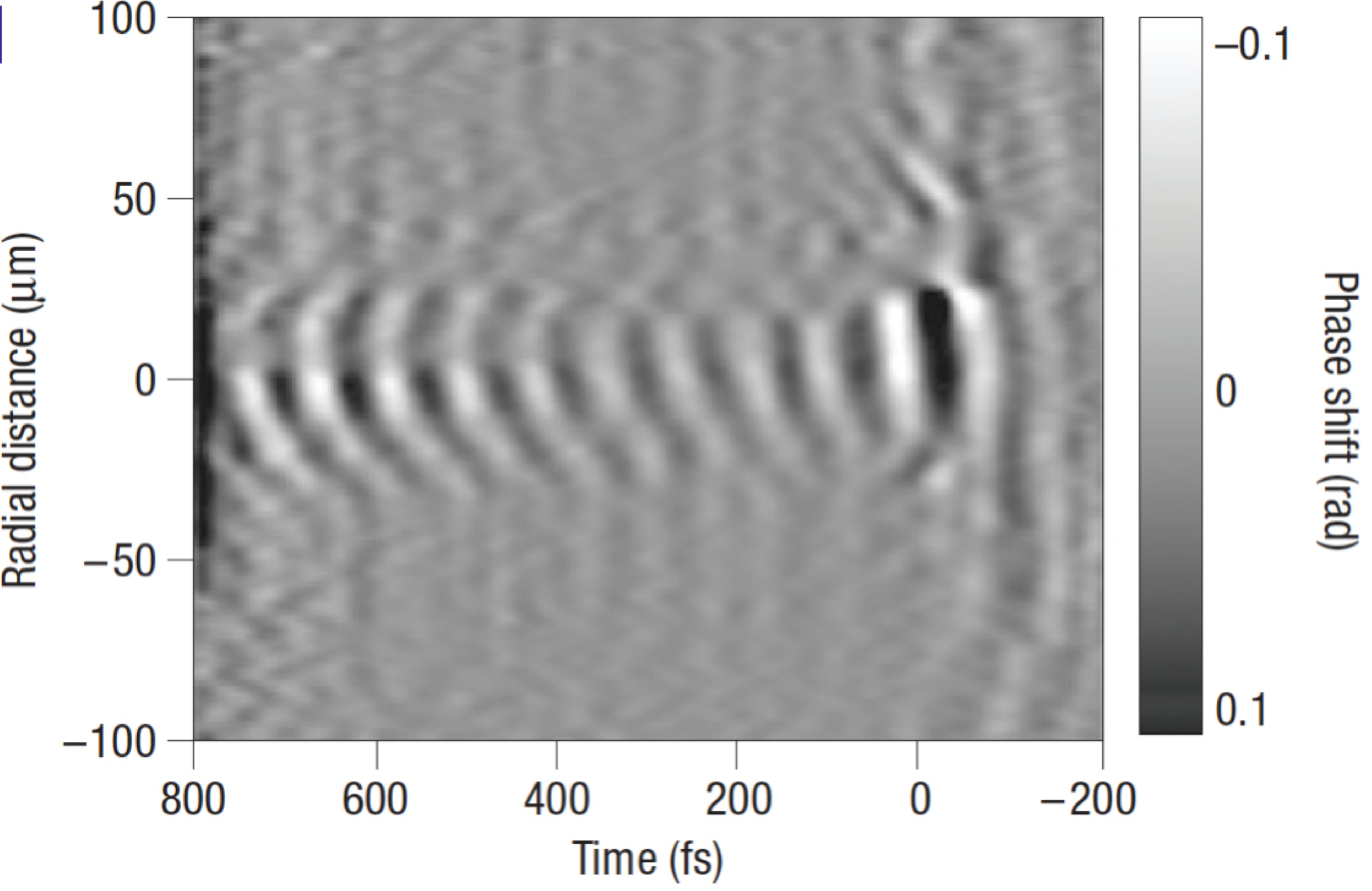}
\caption{The laser pulse, that propagates from left to right, drives a strong wake with relativistic curved front, Courtesy of M. Downer\cite{matl06}.}
\label{fig1} 
\end{figure}

 Since the accelerating field in superconducting Radio-Frequencies cavities is limited to about 100MV/m, the length of accelerators has to increase in order to achieve higher energy gain. To overcome this gigantism issue, J. Dawson \cite{josh07,esar09}) proposed to use a plasma, which, as a ionized medium, can support and sustain extreme electric fields. The pioneer theoretical work performed in 1979 by Tajima and Dawson \cite{taji79} has shown how an intense laser pulse can excite a wake of plasma oscillation through the non linear ponderomotive force associated to the laser pulse. In their proposed scheme, relativistic electrons were injected externally and were accelerated through the very high electric field sustained by relativistic plasma waves driven by lasers. In this former article \cite{taji79}, the authors have proposed two schemes: the laser beat wave and the laser wakefield. Several experiments have been performed in the beginning of the nineties following their idea, and injected electrons at the few MeV level have indeed been accelerated by electric fields in the GV/m range in a plasma medium using either the beat wave or the laser wakefield scheme. Before the advent of short and intense laser pulses, physicists have used the beat wave of two long laser pulses (i. e. with duration much greater than the plasma period) of a few tens of ps with two frequencies $\omega_{1}$ and $\omega_{2}$ to drive the relativistic plasma waves in a perfect homogenous plasma at a density for which the plasma frequency ($\omega_{p}$) satisfies exactly the matching condition, $\omega_{p} = \omega_{1}-\omega_{2}$. The first observation of relativistic plasma waves was performed using Thomson scattering technique by the group of C. Joshi at UCLA \cite{clay85}. Acceleration of 2 MeV injected electrons up to 9 MeV \cite{clay93} and later on, up to 30 MeV \cite{ever94} has been demonstrated by the same group using a CO$_{2}$ laser of about 10 $\mu$m wavelength.  At LULI, 3 MeV electrons have been accelerated up to 3.7 MeV in beat wave experiments with Nd:Glass lasers of about 1$\mu$m wavelengths by a longitudinal electric field of 0.6 GV/m \cite{amir95}. Similar work was also performed in Japan at University of Osaka \cite{kita92}, in U.K. at Rutherford Appleton Laboratory \cite{dyso96}, or in Canada at Chalk River Laboratory \cite{ebra94}. 
Plasma waves driven in the laser wakefield regime at LULI by a few J, 300 fs laser pulse have been used to accelerate 3 MeV injected electrons up to 4.6MeV\cite{amir98}.
In all these experiments, because of the duration of the injected electron which was much longer than the plasma period and even longer than the life time of the plasma, only a very small fraction of injected electrons were accelerated and the output beam had a very poor quality with a maxwellian-like energy distribution. 
Optical observation of radial plasma oscillation has been observed at LOA \cite{marq96,marq97} and in Austin \cite{side96} with a time resolution of less than the pulse duration by using spectroscopy in the time-frequency domain. More recently, at CUOS, using the same technique but with chirped probe laser pulses, a complete visualization of relativistic plasma wave has been performed in a single shot. These results have revealed very interesting features such as the relativistic lengthening of the laser plasma wavelength \cite{matl06} visible on figure \ref{fig1}.
 
 These first experiments have shown acceleration of externally injected electrons. With the development of more powerful lasers, much higher electric fields were achieved, giving the possibility to accelerate efficiently electrons from the plasma itself to higher energies. A major breakthrough, was obtained in 1994 at Rutherford Appleton Laboratory, where relativistic wave breaking limit was reached \cite{mode95}. In this limit, the amplitude of the plasma wave was so large, that copious number of electrons were trapped and accelerated in the laser direction, producing an energetic electron beam. A few hundreds of GV/m electric field was measured. The corresponding mechanism is called the Self Modulated Laser Wake Field (SMLWF) \cite{Andr92,Mora92,Spra92}, an extension of the forward Raman instability \cite{Josh81,Mori94} at relativistic intensities. In those experiments, the electron beam had a maxwellian-like  distribution as it is expected from random injection processes in relativistic plasma waves. This regime has also been reached for instance in the United States at CUOS \cite{umst96} and at NRL \cite{moor97}. However, because of the heating of the plasma by these relatively ``long'' pulses, the wave breaking occurred well before reaching the cold wave breaking limit. The maximum amplitude of the plasma wave in the range 20-60 \% \cite{clay98} was measured using a Thomson scattering diagnostic. Energetic electron beams were produced with a compact laser working at 10Hz at MPQ \cite{gahn99} in the direct laser acceleration scheme(DLA) and at LOA in the SMLWF \cite{malk01}. At LOA, the increase of the electron peak energy when decreasing the electron plasma density have nicely demonstrated that the dominant acceleration mechanism was due to relativistic plasma waves. 
In 2002, another breakthrough was obtained in the forced laser wakefield where a low divergent electron beams with energies up to 200 MeV was obtained with the 1J ``salle jaune" laser at LOA. In this highly non linear regime, the quality of the electron beam was improved by reducing noticeably the interaction between the laser beam and the electron beam. 
To improve fairly the electron beam quality, one has to reduce electrons injection to a very small volume of the phase space. In general, this means that injected electrons must have a duration much shorter that the plasma period, i.e. much less than ten femtoseconds which is difficult to achieve easily today with current accelerator technology. I will show in this review article how physicists have solved this crucial problem by exploring different schemes like the bubble regime, the density gradient injection technique, the space limited ionization approach and the colliding laser pulses scheme. The \textit{fil rouge} of this article is the physic of electron injection. Following this \textit{fil rouge}, I will show that the control of electron injection in a limited space and time region led us to producing stable and very high quality electron beam with a some level of control over the charge, the energy spread, and the electron beam energy.

 \section{Controlling the injection}
 Controlled injection in laser plasma acceleration that lead to high electron beam quality is particularly challenging due to the very small value of the length of the injected bunch that has to be a fraction of plasma wave wavelength , with typical values in the [10-100$\mu$m] range. Doing so, electrons witness the same accelerating field, leading to the acceleration of a monoenergetic and high quality bunch. Electrons can be injected if they are located at the appropriate phase of the wake and/or if they have sufficient initial kinetic energy.
Different schemes have been demonstrated today and allow to control the phase of injected electrons.

 \subsection{Bubble regime} 
In 2002, using 3D PIC simulations, A. Pukhov and J. Meyer-Ter-Vehn have shown the existence of a very promising acceleration regime, called the bubble regime \cite{pukh02}, that leads to the production of a quasi-monoenergetic electron beam. At lower laser intensity, the blow-out regime \cite{lu06}, also allows to obtain such an electron distribution. In those two regimes, the focused laser energy is concentrated in a very small sphere, of radius shorter than the plasma wavelength. The associated ponderomotive force expels radially electrons from the plasma, forming a positively charged cavity behind the laser, and surrounded by a dense region of electrons. As the radially expelled electrons flow along the cavity boundary and collide at the bubble base, transverse breaking occurs \cite{bula97} providing a well localized region of injection in the cavity. 

   \begin{figure}[htbp]
    \begin{center}
     \includegraphics[width=8.5cm]{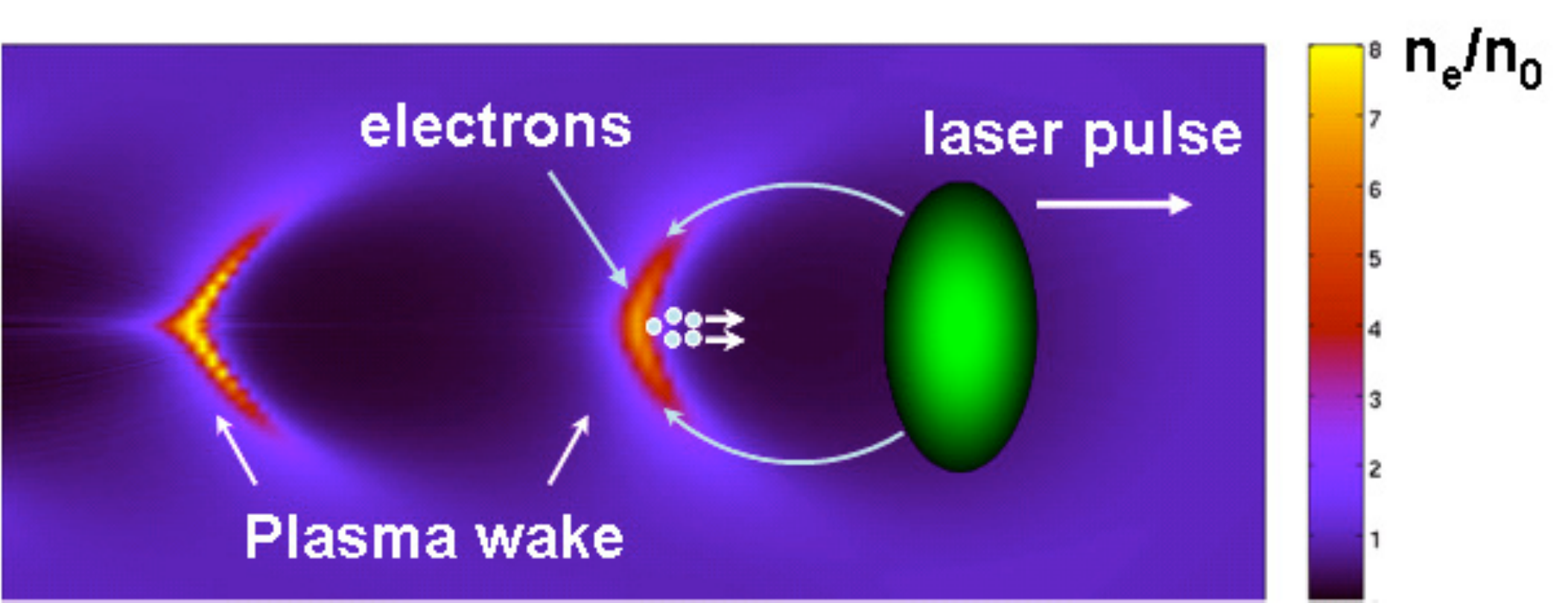}
    \end{center}
    \caption{The laser pulse that propagates from left to right, expels electrons on his path, forming a positively charged cavity. As the radially expelled electrons flow along the cavity boundary and collide at the bubble base, before being accelerated behind the laser pulse.}
   \label{FigRegimeBulle}
   \end{figure}
 
Since the injection is well localized, at the back of the cavity, it gives similar initial properties in the phase space to injected electrons. The trapping stops automatically when the charge contained in the cavity compensates the ionic charge, leading to the generation of a quasi-monoenergetic electron beam that was experimentally demonstrated in 2004\cite{mang04,gedd04,faur04}. Finally, the rotation in the phase-space also leads to a decrease of the spectral width of the electron beam \cite{tsun04}. Electron beam quality is also improved because electrons that are trapped behind the laser do not interacted anymore with the electric field of the laser. The scheme of principle of the bubble/blowout regime is illustrated on figure \ref{FigRegimeBulle}.

\begin{figure}[htbp]
\includegraphics[width=8.5cm]{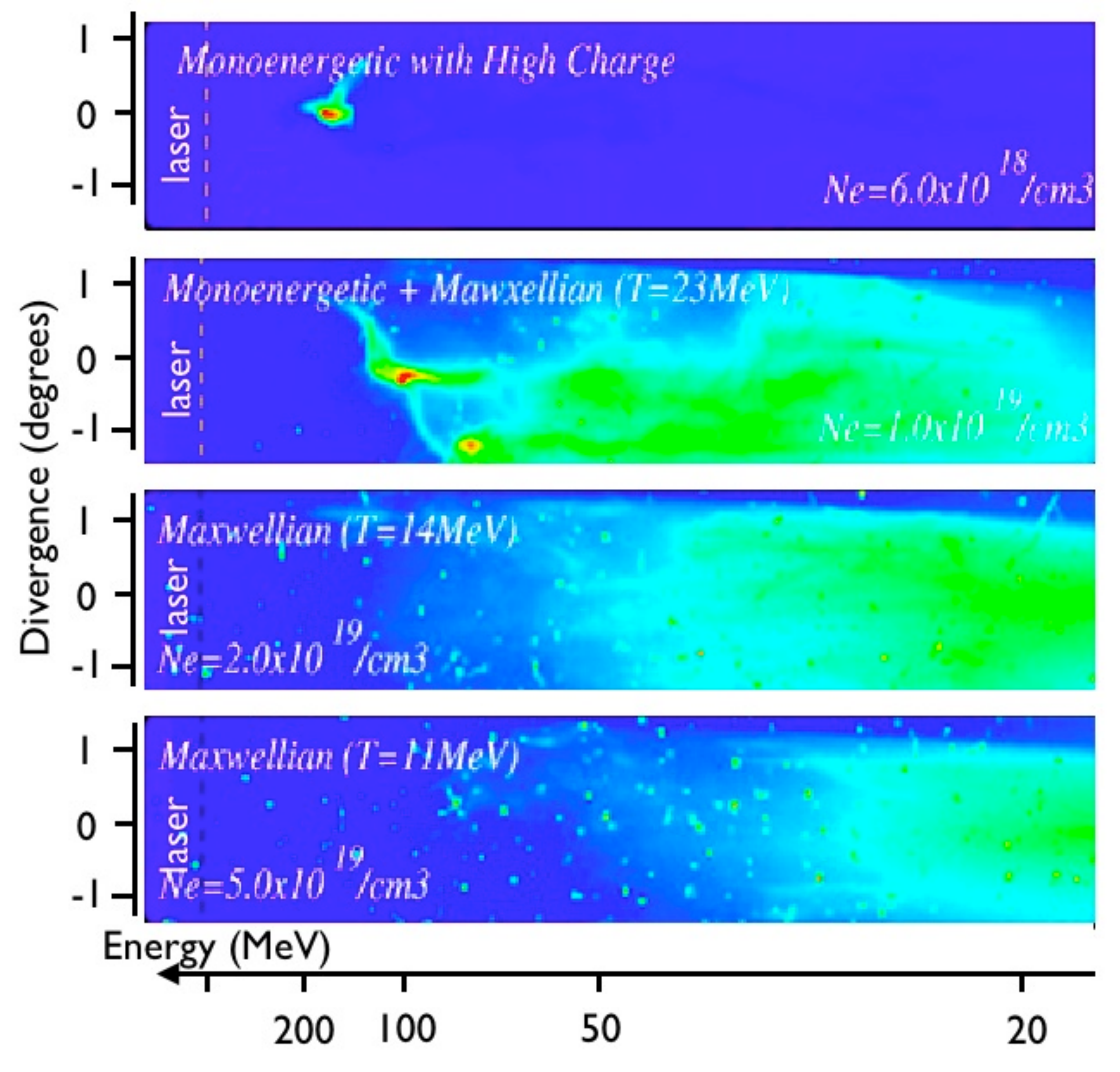}
\caption{\label{fig1bubble} Electron beam distribution for different plasma density showing the transition between the Self Modulated Laser Wakefield, the Forced Laser Wakefield and the Bubble/Blow-out regime. From top to button the plasma density values are $6 \times10^{18}\cmc$, $1 \times10^{19}\cmc$, $2 \times10^{19}\cmc$, and $5 \times10^{19}\cmc$.}
\end{figure}

Several laboratories have obtained quasi monoenergetic electron beams in the bubble/blow-out regime : in France \cite{faur04} with a laser pulse shorter than the plasma period, but also with pulses slightly longer than the plasma period in England \cite{mang04}, in the United States \cite{gedd04}, then in Japan \cite{miur05}, in Tawain \cite{hsie06} and in Germany \cite{hidd06,oste08}, and in Sweden  \cite{mang07}. Electrons at the GeV level were observed in this regime using in a uniform plasma \cite{hafz08,knei09} or in a plasma discharge, i.e, a plasma with a parabolic density profile \cite{leem06} that allows the intense laser beam to propagate over a longer distance, of a few centimeters. In all of those experiments the laser beam parameters did not satisfy fully all the bubble/blow out regime criteria. Nevertheless thanks to the self focusing and self shortening \cite{faur05} effects, the non linear evolution of the laser puls allowed such transverse injection, and some of those experiments are in a regime between the forced laser wakefield and the bubble/blowout regime. The figure \ref{fig1bubble} illustrates in a single experiment the transition between the self modulated laser wakefield, the forced laser wakefield and the bubble/blowout regime. The experiment was performed at Laboratoire d'Optique Appliqu\'ee with the 10 Hz, 30 TW Ti:sapphire laser system operating at wavelength $\lambda_0=0.82\mic$ \cite{pitt02}. The laser pulse that had a duration at Full Width Half Maximum (FWHM) of $\tau=30\fs$ was focused using a 1 m focal length off axis parabolic mirror, onto a focal spot with Full Width Half Maximum (FWHM) of $18\mic$ producing a laser intensity of $3.2\times10^{18}\wcm$. The corresponding normalized vector potential was  $a_0=1.3$. At this intensity, the homogeneous helium gas\cite{malk00,semu01} jet is fully ionized early in the interaction. Since the electron density value is not high, ionization induced refraction did not play an important role. The focal position and its value with respect to the sharp gas jet gradient have been measured and varied in order to optimize the electron beam parameter. This optimum position is found when focusing the laser beam on the edge of the plateau of the gas jet. At high electron plasma densities, for densities greater than $2 \times10^{19}\cmc$, the electron beam has a maxwellian like distribution with a corresponding temperature that increases at lower plasma density. In this range of density the self modulated regime is the dominant acceleration mechanism. At lower density, for densities comprise between $7.5 \times10^{18}\cmc$ and $2 \times10^{19}\cmc$, the forced laser wakefield regime dominates and a plateau appears in the electron beam distribution. The bubble/blowout regime appears only in the very low density range, below  $6 \times10^{18}\cmc$ with a quasi-monoenergetic electron distribution.

It has been shown that with current laser plasma parameters, the bubble/blowout regime was not yet completely established. With the increase of laser power systems, this regime will be reached, and significant improvement of the reproducibility of the electron beam is expected. Nevertheless, since self-injection occurs through transverse wave breaking, it is hardly appropriate for a fine tuning and control of the injected electron bunch. 

 \subsection{Injection in a density gradient}  
One solution to control electron injection with current laser technology has been proposed by S. Bulanov $et$ $al.$ \cite{bula98} using a downward density ramp with a density gradient scale length $L_{grad}$ greater than the plasma wavelength $\lambda_p$.
Injection in a downward density ramp relies on the slowing down of the plasma wave velocity at the density ramp. This decrease of the plasma wave phase velocity lowers the threshold for trapping plasma background electrons and causes wave breaking of the wakefield in the density ramp. This method can therefore trigger wave breaking in a localized spatial region of the plasma. Geddes \emph{et al.} \cite{gedd08} have shown the injection and acceleration of high charge ($>300$ pC) and stable quality beams of $\simeq 0.4\mev$ in the downward density ramp at the exit of a gas jet ($L_{grad}\simeq 100\mic \gg \lambda_p$). These results, although very promising, have the disadvantage that the low energy beam after the plasma blows up very quickly due to space charge effect. To circumvent this issue, one should use a density gradient located early enough along the laser pulse propagation so that electrons can be accelerated to relativistic energies \cite{kim04}. This can be achieved by using a secondary laser pulse to generate a plasma channel transversely to the main pulse propagation axis for instance \cite{chie05}. Doing so, the electron beam energy has been tuned by changing the position of the density gradient. Nevertheless, because of the small value of the laser energy, the electron beam had a large divergence and a Maxwellian energy distribution. 2D PIC simulations have shown that this method can result in high quality quasi monoenergetic electron beams \cite{toma03}. Trapping across a laser generated plasma channel, which has the advantage of ease of production and control has also been considered in simulations \cite{kim04}.\\
	At LOA a density gradient across a laser created plasma channel has been used to stabilize the injection \cite{faur10}. The experiment was performed at an electron density close to the resonant density for the laser wakefield ($c\tau\sim\lambda_p$) to guaranty a post acceleration that deliver high quality electron beams with narrow divergences (4 mrad) and quasi monoenergetic electron distributions with 50 to 100 pC charge and 10\% relative energy spread. 
The use of density gradients at the edges of a plasma channel has shown an improvement of the beam quality and of the reproducibility with respect to those produced in the bubble/blowout regime with the same laser system and with similar laser parameters. However, the electron energy distribution was still found to fluctuate from shot to shot. This injection scheme is promising because it permits to relax the experimental requirements in terms of synchronization (ns) and spatial alignment ($100\mic$). The performances of the experiment could be further improved and could potentially lead to more stable and controllable high quality electron beams. In particular, sharper gradients with $L_{grad}\simeq\lambda_p$ coupled with a long plasma might lead to better beam quality\cite{bran08}.

 \begin{figure}[t]
\includegraphics[width=8.5cm]{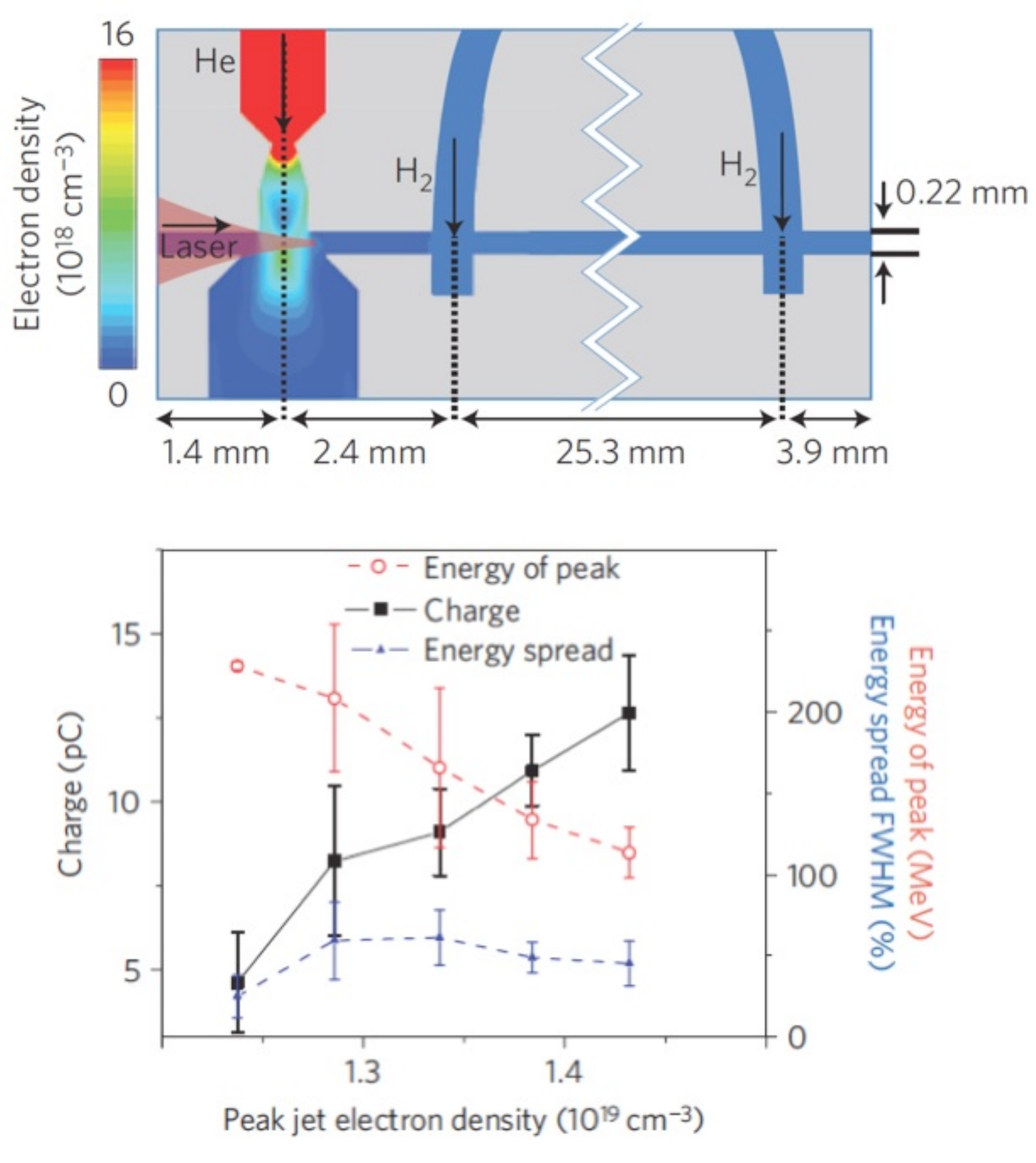}
\caption{\label{fjetcap} Top: The target schematic representation with embedded supersonic gas jet into a capillary that is filled with hydrogen gas, Buttom : the charge (squares), energy (circles), and energy spread (triangles) as a function of the peak jet density. From A. J. Gonsalves \textit{et al.}\cite{gons11}}
\end{figure}

\begin{figure}[t]
\includegraphics[width=8cm]{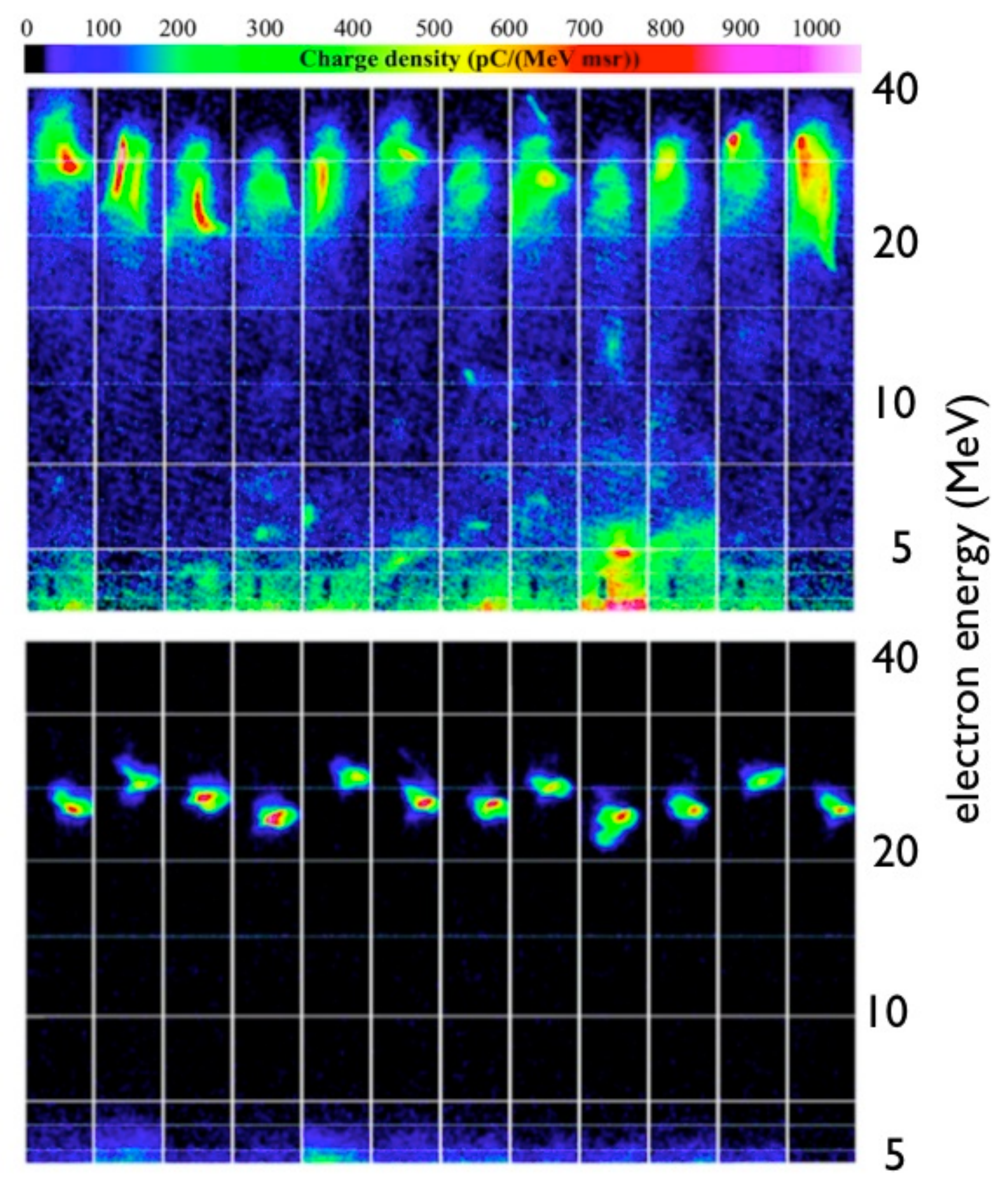}
\caption{\label{fig2gradient} A few shots representative for those 10\% of all the shots with lowest energy spread for self-injection (top) and injection at a density transition (buttom). The horizontal axis in each image corresponds to the transversal electron beam size; the vertical axis shows electron energy. From K. Schmid \textit{et al.}\cite{Schm10}.}
\end{figure}

For example, at LBNL as shown on figure \ref{fjetcap}, coupling the gas jet with a plasma discharge \cite{gons11}, more energetic electrons at 30 MeV has been produced in the density ramp, and have been accelerated up to 400 MeV in a 4 cm parabolic plasma channel. Here also, the density gradient injection has allowed an improvement of the stability and of the electron beam quality. The electron energy, divergence, charge and relative energy spread were found to be respectively 400 MeV, 2 mrad, 10 pC and 11\%.
It has been shown that steeper density transitions, with $L_{grad} \ll \lambda_p$, can also cause trapping\cite{suk01}. Using a shock-front created at the knife-edge of a gas jet such injection has been successfully demonstrated experimentally \cite{Koya09,Schm10}. 
 The figure \ref{fig2gradient} illustrates the improvement of injection in a sharp density gradient, with a characteristic length on the order of the plasma wavelength and a peak electron density of about $5 \times10^{19}\cmc$. The experiment was performed at Max-Planck-Institut fur Quantenoptik using a multi-TW sub-10-fs laser system that delivers for this experiment pulses with 65 mJ energy on target and a duration of 8 fs FWHM. The laser pulse was focused down a spot diameter of $12\mic$ FWHM onto the gas target yielding at a peak intensity of $2.5\times10^{18}\wcm$. The comparison between the self injection and density transition injection shows a reduction of the relative energy spread and of the charge of a about a factor of 2.
 
 \subsection{Injection with colliding laser pulses}  
 
\begin{figure}[t]
\includegraphics[width=8.5cm]{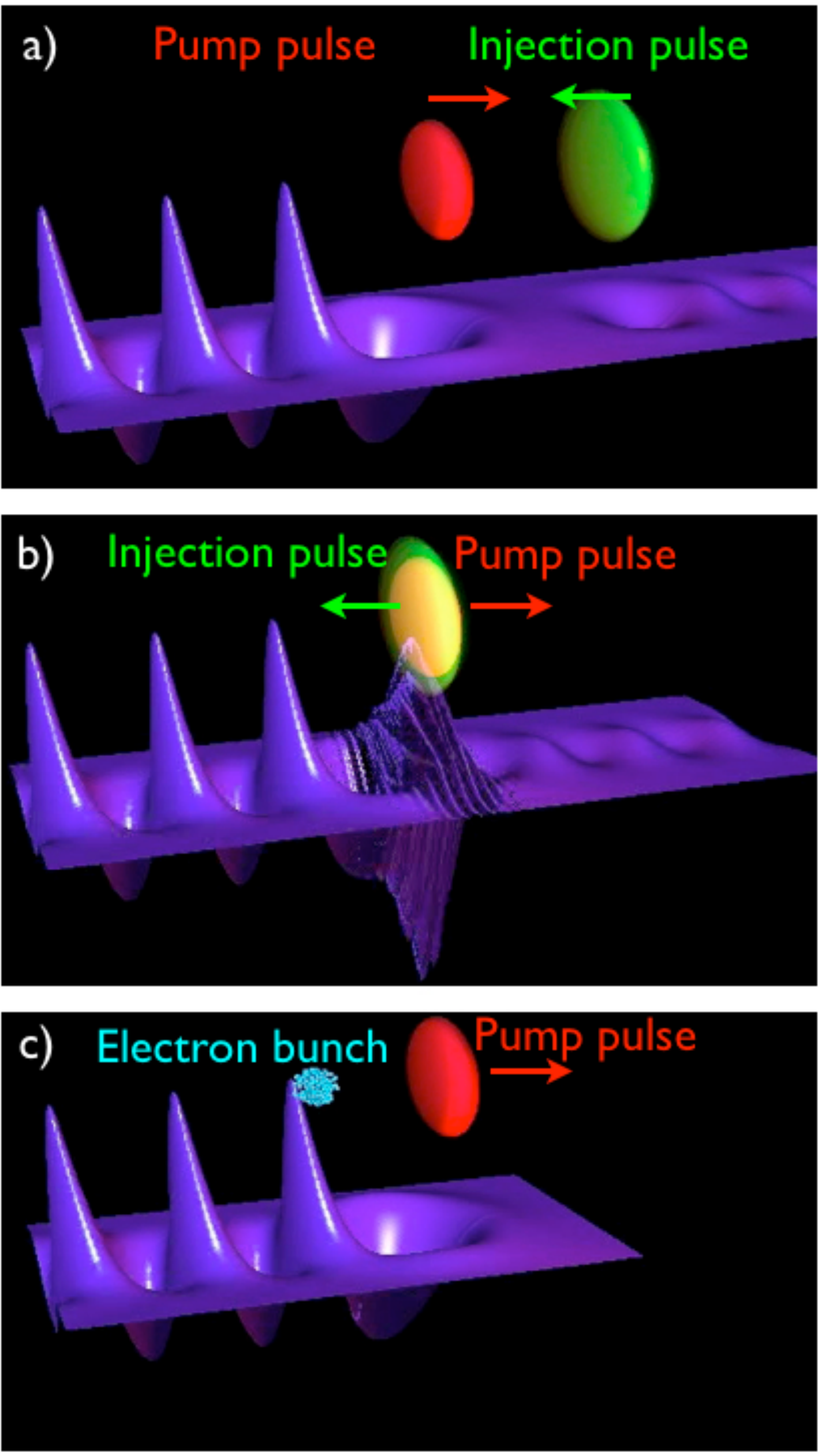}
\caption{\label{fig3colliding_principle} Scheme of principle of the colliding laser pulses: (a) the two laser pulses propagate in opposite direction, (b) during the collision, some electrons get enough longitudinal momentum to be trapped by the relativistic plasma wave driven by the pump beam, c) trapped electrons are then accelerated following the pump laser pulse.}
\end{figure}

 In 2006, stable and tunable quasimonoenergetic electron beams were measured by using two laser beams in the colliding scheme with a counterpropagating geometry. The use of two laser beams instead of one offers more flexibility and enables one to separate the injection from the acceleration process\cite{faur06}. The first laser pulse, the pump pulse, is used to excite the wakefield while the second pulse, the injection pulse, is used to heat electrons during its collision with the pump pulse. After the collision has occurred, electrons are trapped and further accelerated in the wakefield, as shown on figure \ref{fig3colliding_principle}.
 
To trap electrons in a regime where self-trapping does not occur, one has either to inject electrons with energies greater that the trapping energy or dephase electrons with respect to the plasma wave. As mentioned earlier, electrons need to be injected in a very short time ($<\lambda_p/c$) in order to produce a monoenergetic beam. This can be achieved using additional ultrashort laser pulses whose purpose is only restricted to triggering electron injection. Umstadter $et$ $al.$ \cite{umst96}
first proposed to use a second laser pulse propagating perpendicular to the pump laser pulse. The idea was to use the radial ponderomotive kick of the second pulse to inject electrons. Esarey $et$ $al.$ \cite{esar96} proposed a counter-propagating geometry based on the use of three laser pulses. This idea was
further developed by considering the use of two laser pulses \cite{fubi04}. In this scheme, a main pulse (pump pulse)
 creates a high amplitude plasma wave and collides with a secondary pulse of lower intensity. The interference of the two beams creates a beatwave pattern with a zero phase velocity, that heats some electrons from the plasma background. The force associated with this ponderomotive beatwave is inversely proportional to the laser frequency is therefore many times greater than the ponderomotive force associated with the pump laser that is inversely proportional to the pulse duration at resonance. Therefore, the mechanism is still efficient even for modest values of laser intensities. Upon interacting with this field pattern, some background electrons gain enough momentum to be trapped in the main plasma wave and then accelerated to high energies. As the overlapping of the lasers is short in time, the electrons are injected in a very short distance and can be accelerated to an almost monoenergetic beam. This concept has been validated in an experiment \cite{faur06}, using two counter-propagating pulses. Each pulse had a duration of 30 fs at full width half maximum (FWHM), with $a_0=1.3$, $a_1=0.4$. They were propagated in a plasma with electron density $n_e=7 \times 10^{18} cm^{-3}$ corresponding to $\gamma_p=k_0/k_p=15$. It was shown that the collision of the two lasers could lead to the generation of stable quasi-monoenergetic electron beams. The beam energy could be tuned by changing the collision position in the plasma as shown on figure \ref{controlpara}.
 
\begin{figure}[t]
\includegraphics[width=8cm]{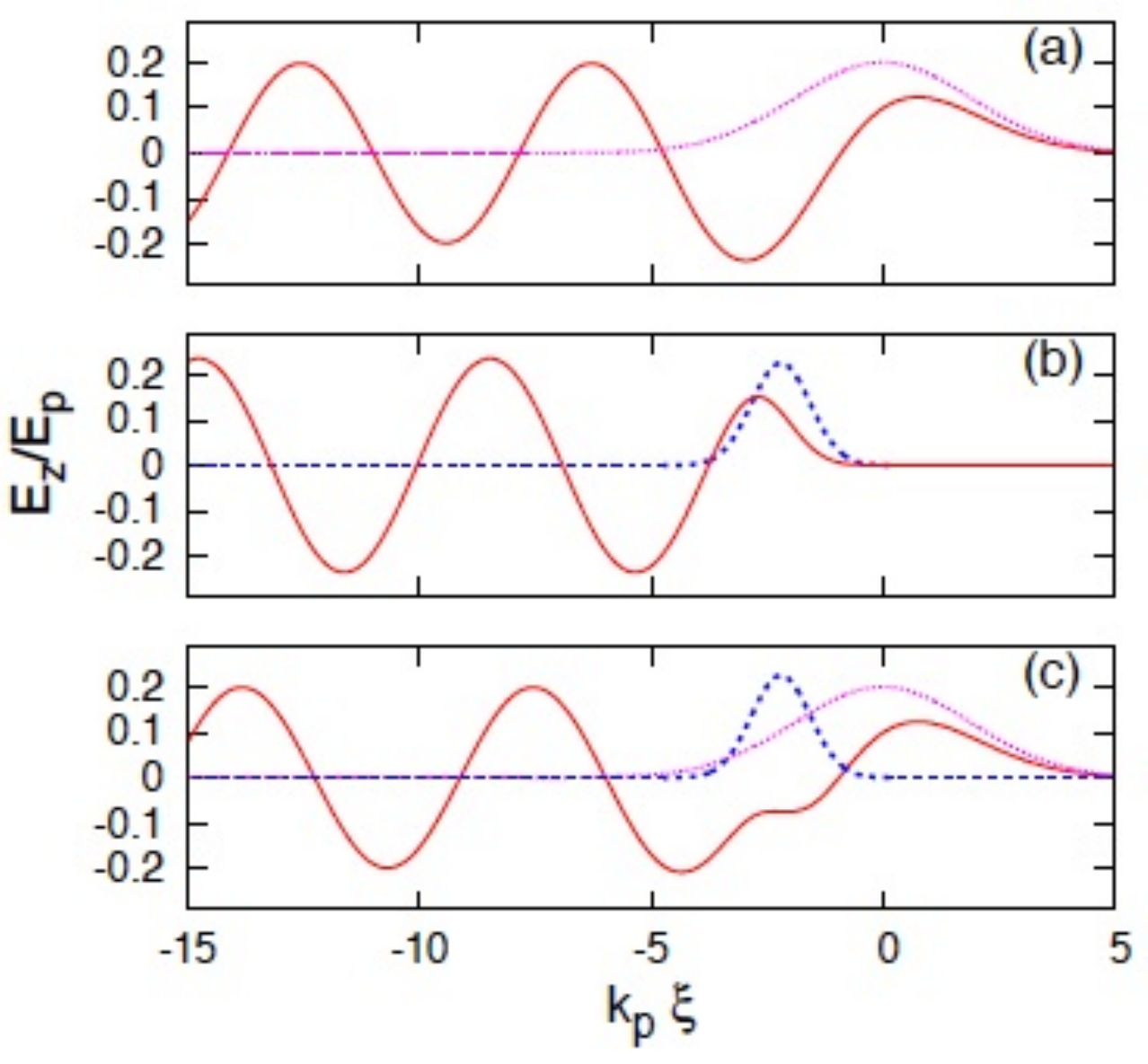}
\caption{\label{beamload} In red, normalized longitudinal electric field. Case a) the laser (in pink) wakefield. Case b) the electron bunch (in blue) wakefield. Case c) resulting fields when laser and electron beams are presented together. Parameters are $a_0=1$, the laser pulse duration 30 fs, $n_e=7 \times 10^{18} cm^{-3}$, $n_{beam}=0.11\times n_e$, bunch duration of 10 fs and diameter 4 microns. From C. Rechatin PhD Thesis.}  
\end{figure}
\begin{figure}[t]
\includegraphics[width=8cm]{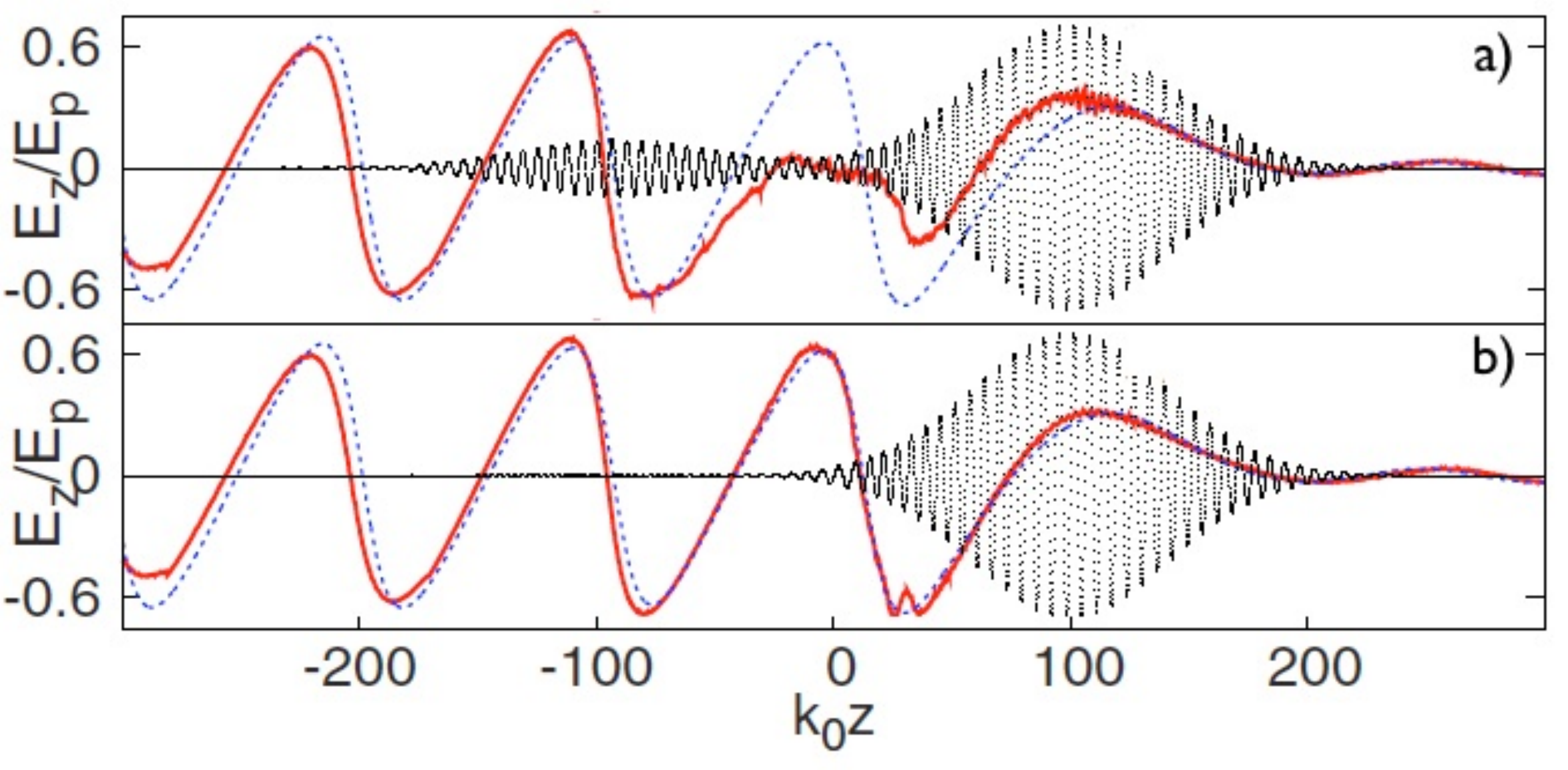}
\caption{\label{diffcollipar_perp} Longitudinal electric field computed at $t=43fs$ in 1D PIC simulation (solid red line), and in fluid simulations (dotted blue line). The transverse electric field is also represented (thin dotted line). Parameters are $a_0=2$, $a_1=0.4$, 30 fs is the laser pulse duration at FWHM with a wavelength of 0.8 micron. The laser pulses propagate in a plasma with electron density $n_e=7 \times 10^{18} cm^{-3}$. In a) the case of parallel polarization and in b) the case of crossed polarization.}  
\end{figure}
\begin{figure}[!h]
\includegraphics[width=8cm]{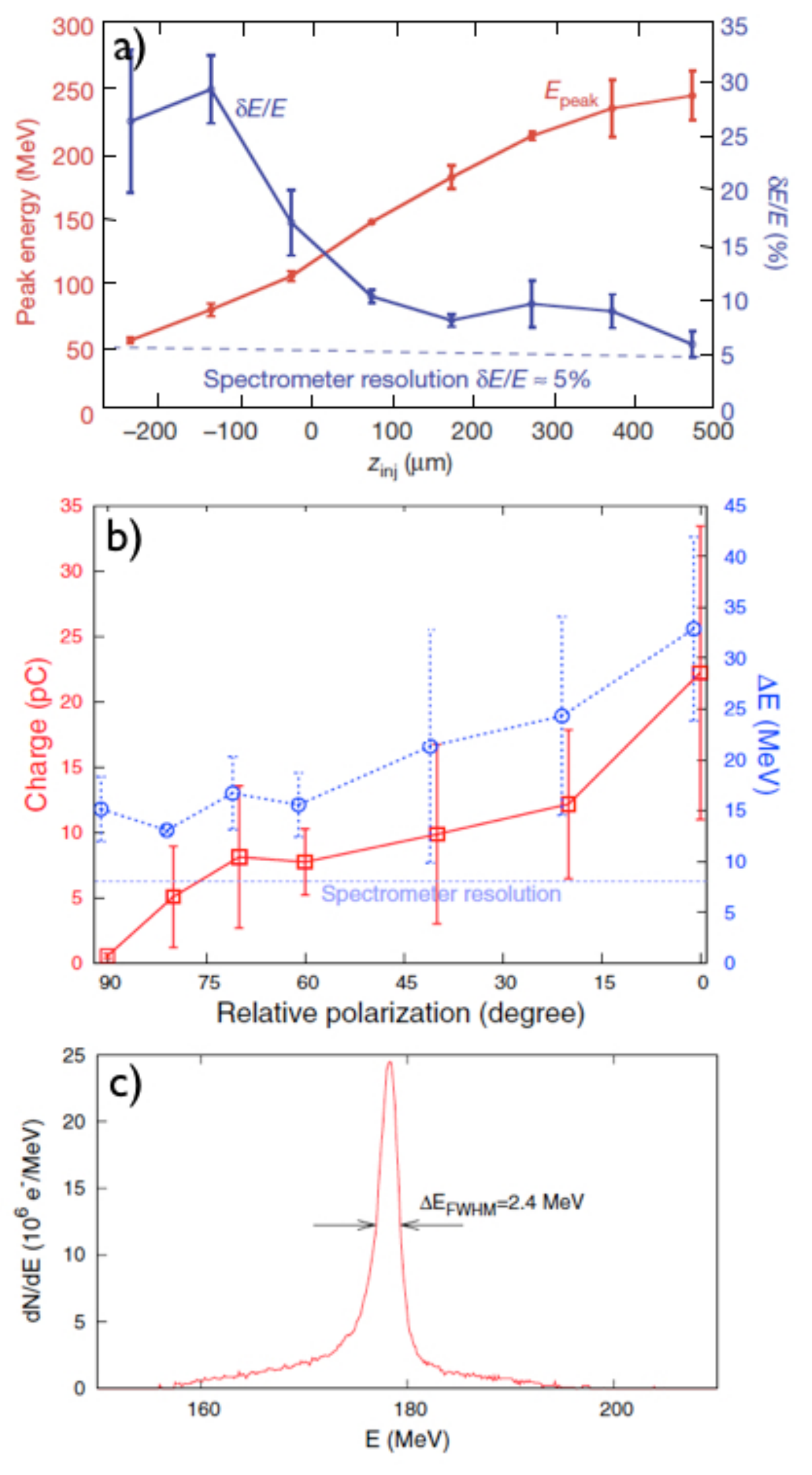} 
\caption{\label{controlpara} a) Evolution of the electron beam peak energy and its energy spread as function of collision position $z_{inj}$ for two parallel polarized laser beams. The electron beam peak energy is shown in red, and the energy spread in blue. Each point is an average of 3–5 shots and the error bars correspond to the standard deviation. The position $z_{inj}=0$ corresponds to injection at the middle of the gas jet, whereas $z_{inj}=500 \mic$ corresponds to early injection close to the entrance of the gas jet, from J. Faure  \textit{et al.} \cite{faur06}. b) Evolution of charge (red solid line with squares), $ {\Delta} E$ at FWHM (blue dotted line with circles) with the angle between the polarizations of injection and pump lasers ( $0{^\circ}$, parallel polarizations; $90{^\circ}$, crossed polarizations). $a_{0}=1.5$, $a_{1}=0.4$, 3 mm gas jet, $n_e=5.7 \times 10^{18} cm^{-3}$, $z_{coll}= -450\mic$. c) deconvolved spectra with a high resolution spectrometer measurement. Physical parameters: $a_{0}=1.2$, $a_{1}=0.35$, 3 mm gas jet, $n_e=7.1 \times 10^{18} cm^{-3}$, $z_{coll}= -300\mic$, from C. Rechatin  \textit{et al.} \cite{rech09}.}
\end{figure}

1D Particle in cell (PIC) simulations have been used to model electron injection in the plasma wave at the collision of the two lasers,
and their subsequent acceleration. The PIC simulations have been compared to existing fluid models \cite{esar96} with prescribed electric field and they show major
differences, such as the plasma fields behavior and the amount of injected charge. The fluid approach fails to describe qualitatively and quantitatively many of the physical mechanisms that occur during and after the laser beams collision \cite{rech07}. In this approach, the electron beam charge has been found to be one order of magnitude greater than the one obtained in PIC simulations. For a correct description of injection, one has to describe properly (i) the heating process, e.g. kinetic effects and their consequences on the dynamics of the plasma wave during the beating of the two laser pulses, (ii) the laser pulse evolution which governs the dynamics of the relativistic plasma waves \cite{davo08}. New unexpected feature have shown that heating mechanism can be achieved when the two laser pulses are crossed polarized. The stochastic heating can be explained by the fact that for high laser intensities, the electron motion becomes relativistic which introduces a longitudinal component through the $\mathbf{v}\times \mathbf{B}$ force. This relativistic coupling makes it possible to heat electrons even in the case of crossed polarized laser pulses \cite{malk09}. Thus, the two perpendicular laser fields couple through the relativistic longitudinal motion of electrons. The heating level is modified by tuning the intensity of the injection laser beam or by changing the relative polarization of the two laser pulses \cite{rech09c}. This consequently changes the volume in the phase space and therefore the charge and the energy spread of the electron beam. Figure \ref{diffcollipar_perp} shows at a given times (42 fs) the longitudinal electric field, during and after collision for parallel and crossed polarization. The solid line corresponds to the PIC simulation results whereas the dotted line corresponds to the fluid calculation. The laser fields are also represented by the thin dotted line. When the pulses have the same polarization, electrons are trapped spatially in the beatwave and can not sustain the collective plasma oscillation inducing a strong inhibition of the plasma wave which persists after the collision. When the polarizations are crossed, the motion of electrons is only slightly disturbed compared to their fluid motion, and the plasma wave is almost unaffected during the collision, which tends to facilitate trapping. 

Importantly it has been shown that the colliding pulse approach allows a control of the electron beam energy which is done simply by changing the delay between the two laser pulses \cite{faur06}. The robustness of this scheme has also allowed to carry out very accurate studies of the dynamic of electric field in presence of high current electron beam.
Indeed, in addition of the wakefield produced by the laser pulse, a high current electron beam can also drive its own wakefield as shown on figure \ref{beamload}. 

This beam loading effect has been used to reduce the relative energy spread of the electron beam. It has been demonstrated that there is an optimal load which flattened the electric field, accelerating all the electrons with the same value of the field, and producing consequently an electron beam with a very small, 1\%, relative energy spread \cite{rech09b}. Thanks to the beam loading effect, the more energetic electrons are slightly slowed down and accelerated at the same energy that the slower one. In case of lower charge, this effect doesn't play any role and the energy spread depends mainly of the heating volume. For higher current, the load is too high and the most energetic electrons slow down too much and get energies even smaller that the slower one \cite{rech09b}, increasing the relative energy spread. The optimal load was observed experimentally and supported by full 3D PIC simulations, its corresponds to a peak current in the 20-40 kA range. The decelerating electric field due to the electron beam was found to be in the GV/m/pC range. 

\subsection{Injection triggered by ionization}
\begin{figure}[t]
\includegraphics[width=8cm]{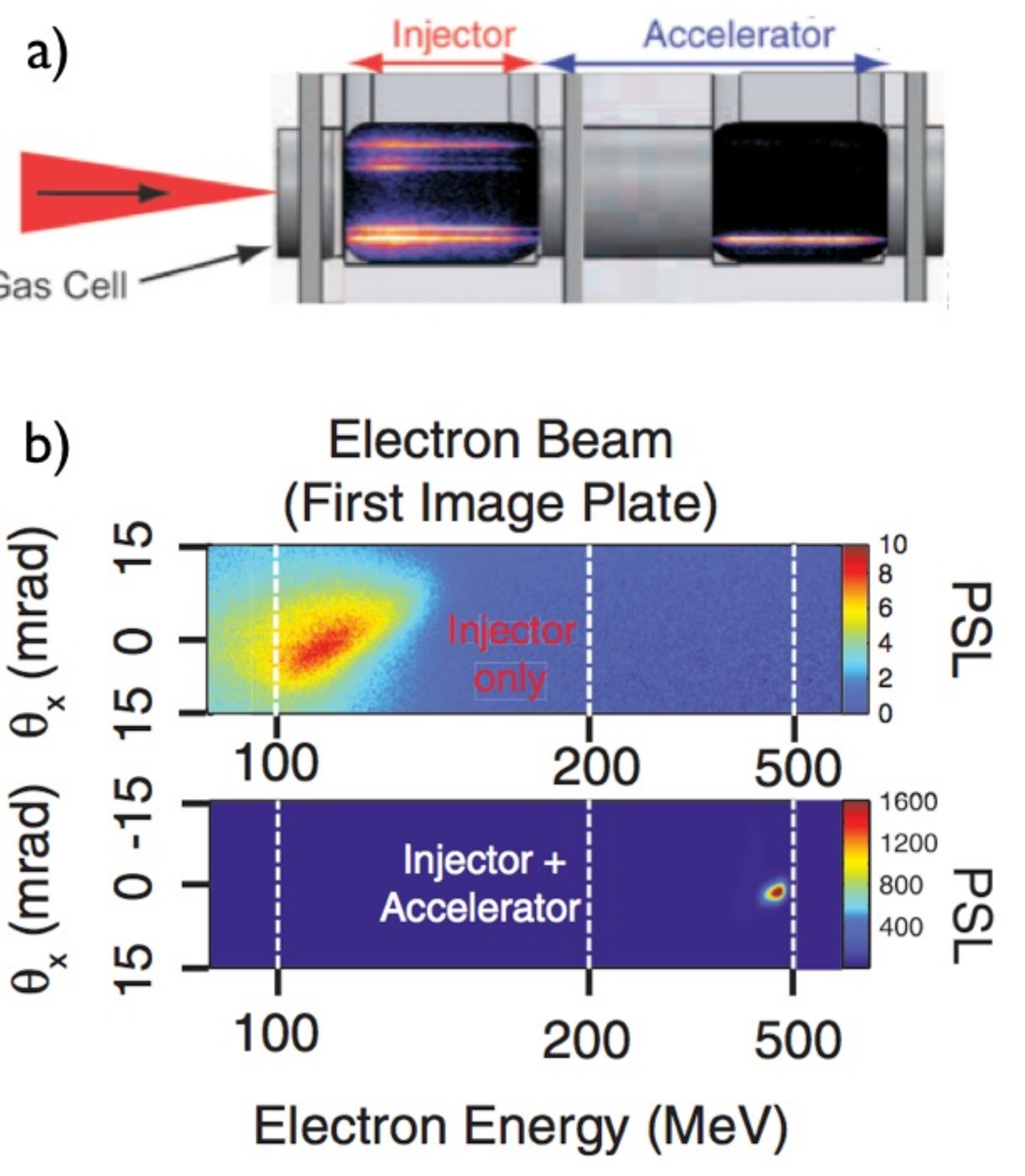}
\caption{\label{twostagel} a) Schematic of the experimental setup showing the laser beam, the two-stage gas cell, on left the injector part and on right the accelerator part. b) Magnetically dispersed electron beam images from a 4 mm injector-only gas cell (top) and the 8 mm two-stage cell (bottom). From B. B. Pollock  \textit{et al.} \cite{Clay11}.}  
\end{figure}

Another scheme has been proposed recently to control the injection by using a high Z gas and/or a high Z-low Z gas mixture. Thanks to the large difference in ionization potentials between successive ionization states of trace atoms, with a single laser pulse, one can drive relativistic plasma waves in the leading edge of the laser pulse that ionizes easily the low energy level atoms, and when the laser intensity is close to its maximum ionizes the more internal level. Such related ionization trapping mechanism has been first demonstrated in electron beam driven plasma wave experiments on the Stanford Linear Collider (SLAC) \cite{oz07}. Trapped electrons from ionization of high Z wall ions from capillary walls was also inferred in experiments on laser wakefield acceleration \cite{Rowl08}. In the case of self guided laser driven wakefield, a mixture of helium and trace amounts of different gases was used\cite{Pak10,Guff10}. In one of these experiments, electrons from the K shell of nitrogen were tunnel ionized near the peak of the laser pulse and were injected into and trapped by the wake created by electrons from majority helium atoms and the L shell of nitrogen. Because of the relativistic self focusing effect, the laser is propagating over a long distance with peak intensity variations that can produce this injection over a too long distance and in a non homogeneous way that does not allow to produce a low relative energy spread electron beam. Importantly, it has been shown that the required laser energy to trap electrons is reduced, rending this approach of great interest to produce electron beams with a large charge at moderate laser energy. To overcome the problem of delocalized injection over a long distance, as shown on figure \ref{twostagel}, experiments using two gas cells have been performed at LLNL\cite{Clay11}. By restricting electron injection to a distinct short region, in the first short cell filled with gas mixture (the injector stage), energetic electron beams (of the order of 100 MeV) with a relatively large energy spread have been generated. Some of these electrons are then further accelerated by a second, longer accelerator stage composed by a longer cell filled with low-Z gas, which increases their energy to 0.5 GeV while reducing the relative energy spread to $<5 \% $ FWHM.

\section{Future of the laser plasma accelerators}
\label{sec:3}
The work achieved this last decade on the development of laser plasma accelerator has been paved by many successful (and unsuccessful) experiments. Thanks to these pioneering works and judging from the incredible results achieved recently, the time has come where technological issues have to be addressed if one wants to build a reliable machine that will operate in a robust way.
 The recent diagnostics that have been developed and/or adapted to measure the laser, plasma and electron beam parameters have played a major role in understanding relativistic laser plasma interaction. These accurate measurements are crucial for performing numerical simulations with input parameters as close as possible to the experimental ones. In some case, one to one simulations allows to reproduce the measured electron beam parameters using 3D PIC simulations without having any adjustable parameters \cite{davo08}.
 The exploration of new numerical schemes is important to perform accurate simulations nearly free from numerical noises to describe higher quality electron beams with lower emittance values and with lower relative energy spread values. These improvements should allow for example to validate theoretical work on two stages laser plasma design that should allow the development of few GeV electron beams \cite{malk06} with a small relative energy spread and a good emittance and to prepare relevant diagnostics. In parallel, theoretical and experimental research should of course be pursued to explore new regimes and to validate theories and numerical codes.
 
\begin{figure}[!h]
\includegraphics[width=7cm]{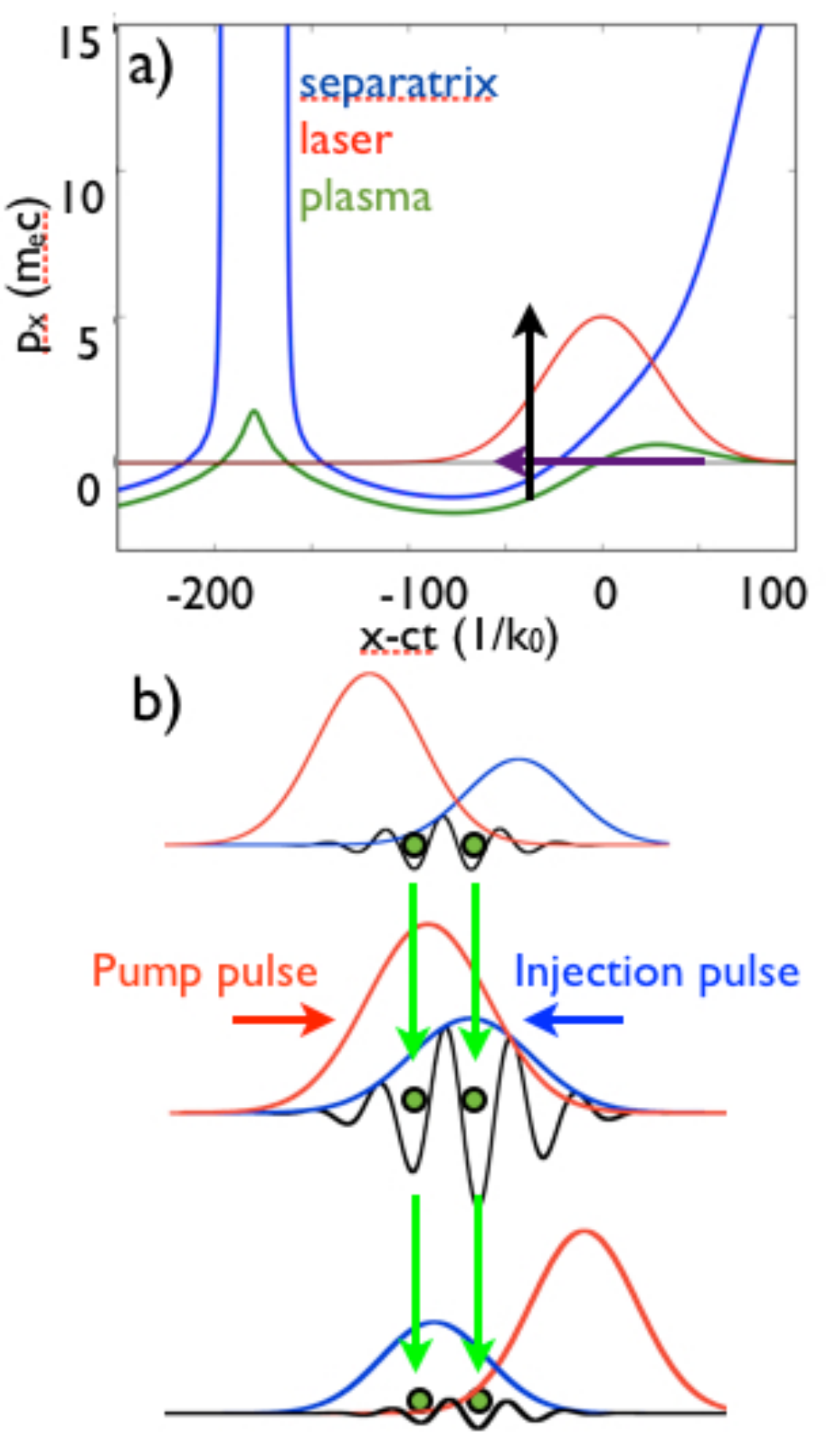} 
\caption{\label{hotcoldl} a) injection principle in the colliding laser pulses: in the ``hot" injection scheme, injection is achieved thanks to momentum gain (black arrow) that electrons from the plasma wave (green curve) reach during the collision that allow them to cross the separatrix (blue curve). Injection principle in the cold injection scheme: electrons are injected by being dephased from the front of the main pulse to its back without momentum gain (purple arrow). After dephasing, electrons are effectively injected over the separatrix. b) Principle of electrons dephasing in a standing wave generated by the collision with a counter-propagating circular laser pulses. From X. Davoine  \textit{et al.} \cite{davo10}.}  
\end{figure}

New ideas regarding injection control, such as the cold injection scheme or assisted magnetic field scheme, are always welcome for future possible electron beam quality improvements. In the cold injection scheme \cite{davo10}, two laser pulses with circular polarization collide in the plasma and produce a standing wave that freeze electrons at the collision point in a standing wave as shown on figure \ref{hotcoldl}. After the collision, these electrons are accelerate in the plasma wave that is restored thank to the wakefield generate by the pump beam. In this case, no heating is needed and electrons cross the separatrix because they keep a constant longitudinal momentum. The use of an external magnetic field was recently proposed\cite{viei11} to control off-axis injection bursts in laser driven plasma waves. It has been shown theoretically that this magnetic field can relax the injection threshold and can be used to control the main output beam features such as charge, energy, and transverse dynamics in the ion channel associated with the plasma blowout. 
 
 In the near future, the development of compact free electron lasers that could deliver an intense X ray beam in a compact way by coupling the electron beam with undulators. Thanks to the very high peak current of a few kA\cite{Lund11} comparable to the current used at LCLS, the use of laser plasma accelerators for free electron laser, the so-called fifth generation light source, is clearly identified by the scientific community as a major development. For these applications, one has to reduce the relative energy spread and to solve the problem of beam transport between the electron beam and the undulator by preserving the parameters of the electron beam. Due to the large divergence of the electron beam that is proper of laser plasma accelerators, one has to use for example ultra high magnetic gradient quadrupoles to reduce the temporal stretching of electrons that have a longer path that the on axis electrons. Undulators radiation \cite{fuch09} and synchrotron\cite{schl08} radiation have been recently obtained by coupling an undulator with an electron beam from a laser plasma accelerator.
 In the near term future, alternative schemes to produce ultra short X ray beams, are also considered, such as Compton, betatron or Bremsstralhung X rays sources. Incredible progress have been made on betatron radiation in a laser plasma accelerators, from its first observation in 2004 \cite{rous04} and its first electron betatronic motion observation \cite{glin08}, number of articles have reported in more details on this new source, such as its sub ps duration \cite{taph07} and its transverse size in the micrometer range \cite{taph06}. The betatron radiation has been used recently to perform high spatial resolution, of about 10 microns, X ray contrast phase images in a single shot mode operation \cite{four11,knei11}. This radiation has allowed physicists to determine very subtile informations of electrons injection in capillaries \cite{geno11} and electrons dynamic that cannot be obtained with other diagnostic \cite{cord11}. Measuring the angular and the energy spectra of this radiation gives an alternative method for estimating the transverse electron beam emittance and for confirming previous measurements using the pepper pot technique. The typical value of the normalized transverse emittance is found to be in the few $\pi$.mm.mrad for tens MeV electrons \cite{frit04,sear10,brun10}. The emittance values have been found to increase for lower electron beam energy values \cite{frit04}. This approach for measuring the emittance is particularly pertinent when the pepper pot technique stops to be relevant, for example for higher electron energies. Measuring in the betatron angular distribution the correlation between the ellipticity of the electron beam and the laser polarization, interaction of the accelerated electron bunch with the laser field \cite{mang06b} has been demonstrated, and has also probably been confirmed recently \cite{cipi11}. For the longitudinal beam emittance, Coherent Transition Diagnostics have been used to measure the shortest electron bunch of 1.5 fs RMS. Time resolved magnetic field measurements have confirmed the shortness of the electron bunch produced in laser plasma accelerators \cite{buck11} and have revealed interesting features of laser plasma accelerating and focusing fields \cite{buck11,kalu10}. Optical transition radiation diagnostics has been very useful to identify the existence of one or two \cite{glin07}, or even more electron bunches produced at different arches of the plasma wakefield, and the possible interaction between the electron bunch and the laser field \cite{glin07}.
   The improvement of the laser plasma interaction with the evolution of short-pulse laser technology, a field in rapid progress, will still improve this new and very promising approach which potential societal applications in material science for example for high resolution gamma radiography \cite{glin05, beni11}, for medicine for cancer treatment \cite{glin06b, fuch09b}, chemistry \cite{broz05, gaud10} and radiobiology \cite{malk10, riga10, malk08}.  
  For longer term future, the ultimate goal which is of major interest for high energy physics will require very high luminosity electron and positron beams having TeV energies. To reach these parameters with laser plasma accelerators will take at least 5 decades and significant further works to develop this technology is required. The incredible improvement in the energy and beam quality of laser based plasma accelerators seems promising for high energy physics purposes. But electron energy is not the only important parameter and it is also necessary to consider the extremely high luminosity value required for this objective that has to be, for TeV beams, must be greater than $10^{34}cm^{2}s^{-1}$. Reaching this value will require at least to produce electron bunches at kHz repetition rates with 1 TeV in energy and with 1nC per bunch. The corresponding average power of the electron beam will be of about 1 MW. Assuming in the best case a coupling of 10 \% from the laser to the electron beam (today in the best case this value rise 10 \% (respectively 1\%) for a 10\% (respectively 1\%) relative energy spread electron beam) one has to produce at least 10 MW (respectively 100MW) of photons. Since the laser wall-plug efficiency is below 1\%, one needs at least in the most favorable case 10GW of electrical power to reach this goal. The laser efficiency conversion could be increased up to 50\% by using diode pumped systems, thus reducing the needed power to 0.2 GW. These considerations were done neglecting several other issues such as the propagation of electron beams into a plasma medium, laser plasma coupling problems, laser depletion, emittance requirements and others \cite{krus09}. \\
  Nevertheless, before reaching an objective and more accurate conclusion on the relevance of the laser plasma approach for high energy physics, it will be necessary to design a prototype machine (including several modules) in coordination with accelerator physicists. An estimation of the cost and an identification of all the technical problems that are to be solved will permit an estimate of the risk with respect to other approaches (particle beam interaction in plasma medium, hot or cold technology, or others).
In conclusion while a significant amount of work remains to be done to deliver beams of interest for high energy physics, the control of the electron beam parameters is now achieved and many of the promised applications are become a reality.

 \section*{Acknowledgements}
 
I acknowledge warmly my former PhD students X. Davoine, J. Faure, S. Fritzler, Y. Glinec, C. Rechatin,  and former post-docs J. Lim, A. Lifschitz, O. Lundh, and B. Prithviraj, my co-worker I. Ben-Ismail, S. Corde, E. Lefebvre, A. Rousse, A. Specka, K. Ta Phuoc, and C. Thaury, who have largely contributed during this last decade to the progress done in laser plasma accelerators research at LOA.
I acknowledge the different teams from APRI, CUOS, IC, JAEA, LLNL, LLC, LBNL, MPQ, RAL, UCLA and others, that have in a competitive and fair atmosphere made significant progresses sharing with passion this wonderful adventure.
I also acknowledge the support of the European Research Council for funding the PARIS ERC project (contract number 226424).
%
%



\end{document}